\documentclass[10 pt]{article}

\usepackage{graphicx}
\usepackage{epsfig}
\usepackage{color}
\usepackage{cite}
\usepackage{amssymb,amsmath}
\usepackage{graphicx}  

\newcommand{\beq}{\begin{equation}}
\newcommand{\eeq}{\end{equation}}
\newcommand{\be}{\begin{equation}}
\newcommand{\ee}{\end{equation}}
\newcommand{\bea}{\begin{eqnarray}}
\newcommand{\eea}{\end{eqnarray}}
\newcommand{\ba}{\begin{array}}
\newcommand{\ea}{\end{array}}

\newcommand{\Ko}{{K}^0}
\newcommand{\Kob}{\bar{K}^0}
\newcommand{\kappao}{{\kappa}^0}
\newcommand{\kappab}{\bar{\kappa}^0}
\newcommand{\da}{^\dagger} 
\newcommand{\tr}{{\rm Tr}} 
\newcommand{\cL}{{\cal L}}
\newcommand{\cO}{{\cal O}}

\fontencoding{T1}
\fontfamily{garamond}
\fontseries{m}
\fontshape{it}
\fontsize{13}{15}
\selectfont

\begin{document}

\title{A Theory of Scalar Mesons}
\author{G.~'t Hooft$^a$, G.~Isidori$^b$, L.~Maiani$^{c,d}$,  A.D.~Polosa$^d$, V.~Riquer$^d$,  \\
\\
$^a$ {\it Institute for Theoretical Physics, Utrecht University,}\\ 
 {\it and Spinoza Institute, Postbus 8000, 3508 TA Utrecht, The Netherlands}\\
$^b$ {\it Scuola Normale Superiore, Piazza dei Cavalieri 7, 56126 Pisa, Italy} \\
 {\it and INFN, Laboratori Nazionali di Frascati, Via E.Fermi 40, 00044 Frascati, Italy} \\
$^c$ {\it Dip. di Fisica, Universit\`a di Roma ``La Sapienza'', 
 P.le A. Moro 2, 00185 Roma, Italy}  \\
$^d${\it INFN, Sezione di Roma ``La Sapienza'', P.le A. Moro 2, 00185 Roma, Italy} \\
}
\date{\today}
\maketitle
We discuss the effect of the instanton induced, six-fermion effective Lagrangian on 
the decays 
of the lightest scalar mesons in the diquark--antidiquark picture. 
This addition allows for a remarkably good description of light scalar meson decays. 
The same effective Lagrangian produces a 
mixing of the lightest scalars with the positive parity $q{\bar q}$ states. Comparing with 
previous work where the $q{\bar q}$ mesons are identified with the nonet at 1200-1700 MeV, 
we find that the mixing required to fit the mass spectrum is in good agreement with the 
instanton coupling obtained from light scalar decays. 
A coherent picture of scalar mesons as a mixture of tetraquark states 
(dominating in the lightest mesons) and heavy $q{\bar q}$ states
(dominating in the heavier mesons) emerges.

\section{Introduction}
We study in this paper the strong decays of the lightest scalar mesons: 
$\sigma$, $\kappa$, $f_0$, $a_0$ and the relations of the light scalars to 
the scalar mesons observed in the 1-2 GeV range. 

Recent experimental and theoretical evidence for the existence of  
$\sigma$ and $\kappa$~\cite{scalarlow,BES1,Caprini,Descotes} 
(see also~\cite{previous})  indicates that light scalars make a full SU(3) flavor nonet. 
Their mass spectrum, with the peculiar inversion of the $\kappa$ and $f_0$ or $a_0$ 
mass ordering, speaks however against the naive $q\bar q$ picture.
The most natural explanation for such complete multiplet with inverted mass spectrum
is that these mesons are diquark--antidiquark bound states.  
The $K \bar K$ molecular constitution~\cite{molecule}, advocated to explain the 
degeneracy of $f_0/a_0$ with the $K \bar K$ threshold, would lead most likely to 
incomplete multiplets. 

The picture where the light scalar mesons are diquark--antidiquark 
states bound by color forces has been discussed by several 
authors~\cite{jaffeetal,Schect,scalar1}.
In this picture the diquarks, which we will indicate with $[q_1 q_2]$, 
are in color ${\bf \bar 3}$, spin $S=0$ and flavor ${\bf \bar 3}$, 
and antidiquarks in the conjugate representations. 
Diquark--antidiquark bound states (tetraquarks, for short) naturally 
reproduce the SU(3) nonet structure with the correct mass ordering, 
as indicated by the explicit quark composition:
\bea
&&\sigma^{[0]}\;=\;[ud][\bar u \bar d]\notag \\
&&\kappa\;=\;[su][\bar u \bar d]; \;[sd][\bar u \bar d] ~{\rm (+~conjugate~doublet)}\notag \\
&&f_0^{[0]}=\frac{[su][\bar s \bar u]+[sd][\bar s \bar d]}{\sqrt{2}}\notag \\
&& a_0=[su][\bar s \bar d];~\frac{[su][\bar s \bar u]-[sd][\bar s \bar d]}{\sqrt{2}};
~[sd][\bar s \bar u]
\label{qcomp}
\eea

While the mass spectrum of the light scalar mesons is well understood 
in terms of diquark--antidiquark bound states, the overall picture 
of scalar mesons is still affected by two drawbacks. First, the 
strong decays into two pseudoscalar mesons ($S\to PP$) 
have so far escaped a satisfactory theoretical understanding 
in the quark rearrangement picture.
In particular, the $f_0 \to \pi\pi$ coupling is too small 
compared to experiments (according to the ideal-mixing decomposition 
in (\ref{qcomp}) it should vanish)
and the $a_0 \to \eta\pi$ coupling largely exceeds its experimental value.
Second, the identification of the $q\bar q$ scalar sates is an open issue. 
The latter must exist, as indicated by the well identified nonets corresponding to\footnote{The 
fact that the $J^{PC}=0^{++}$ components of the  $q\bar q$ states must have $L=1$ can most 
easily be understood as follows. The operator $\bar q(0) q(0)$ can be written as 
$\bar q_L q_R + \bar q_R q_L$. Now $\bar q_L$ and $q_R$ both have helicity $R$ but 
opposite momentum, hence opposite spin. Taking the momenta in the $z$-direction, we 
find that the spin structure of the fields here is  $(1, 2)+(2, 1)$, so $S^{\rm tot}_z=0$. 
This is to be compared with the pion case, where $\gamma_5$ flips the relative sign:  
$\bar q_L q_R - \bar q_R q_L = (1, 2)-(2, 1)$.  The latter is clearly the mode with total 
spin $S=0$, since it can be written as $\epsilon_{ij}  u_i v_j$, but in the scalar case, 
it is the $S_z=0$ component of the state with total spin $S=1$. To obtain total angular
momentum $J=0$ we therefore need orbital angular momentum $L=1$. The gamma algebra then 
shows that such $L=1$ contribution of the wave functions to $\bar q(0)q(0)$ originates from 
the relativistic terms that mix the large components of the Dirac spinors with the small ones.} 
orbital angular momentum $L=1$ and quantum numbers $J^{PC}=1^{++}, 1^{+-}, 2^{++}$~\cite{BorGa}: 
all predicted states are unambiguously identified but for $J^{PC}=0^{++}$.

In this paper we show that these two main problems are solved by 
the instanton induced effective six-fermion Lagrangian~\cite{Gerardo},
the same effective interaction which solves the problem of the 
$\eta-\eta^\prime$ masses~\cite{etapuzzle,eta2}.
Such Lagrangian has two important effects in scalar mesons dynamics: 
(i) it generates a mixing between tetraquarks and $q\bar q$ states,
(ii) it provides an additional amplitude which brings the 
strong decays of the light scalars in good agreement 
with data. Former studies of instanton-induced effects 
in scalar meson dynamics can be found in the literature~\cite{previnst},
but these two effects have not been discussed before.

The two effects induced by the effective instanton Lagrangian
are closely connected. The tetraquark--$q\bar q$ mixing makes it possible
to identify the heavy scalars around $1.5$ GeV as 
predominantly $q\bar q$ states with a non-negligible 
tetraquark component. The latter is essential to explain 
the anomalous mass spectrum of such mesons, 
as originally proposed in~\cite{Schect}. 
On the other hand, integrating out the heavy $q\bar q$ components,
the tetraquark--$q\bar q$ mixing manifests itself into the  
non-standard $S\to PP$ decay amplitude for the light scalar mesons
which improves substantially the agreement with data.
These two independent phenomena lead  to obtain two 
independent phenomenological determinations of the 
non-perturbative parameter which controls the 
matrix elements of the instanton Lagrangian
in the scalar sector. The two determinations turn 
out to agree, reinforcing the 
overall consistency of the picture. 

The paper is organized as follows: in Section~\ref{sect:main} we 
illustrate the two main effects of the instanton interaction in 
the scalar sector, constructing the corresponding effective 
Lagrangians in terms of meson fields. In Section~\ref{sect:SPP} 
we present a numerical analysis of $S\to PP$ decays
in this scheme, demonstrating the relevant  phenomenological
role of the instanton contribution; as a further cross-check, 
we also show that $S\to PP$  decays are badly described under 
a pure $q\bar q$ picture  of 
the light scalar mesons, with or without instantons. 
The results are summarized in the Conclusions.

\section{Instanton effects in scalar meson dynamics}
\label{sect:main}

QCD instantons produce an effective interaction which reduces 
the ${\rm U}(N_f)_L \times {\rm U}(N_f)_R$ global symmetry of the quark model 
in the chiral limit to ${\rm SU}(N_f)_L \times {\rm SU}(N_f)_R$ times baryon number. 
The effect can be described by the following effective 
Lagrangian~\cite{etapuzzle} (see also~\cite{shifman}):
\be
\cL_{\rm I} \propto  {\rm Det} (Q_{LR})~, \qquad  (Q_{LR})^{ij} = {\bar q}^i_{L} q_R^j~,
\ee
where $i$ and $j$ denote flavor indices (summation over color indices is understood).

\begin{figure}[t]
\begin{center}
\epsfig{height=3truecm, width=12truecm,figure=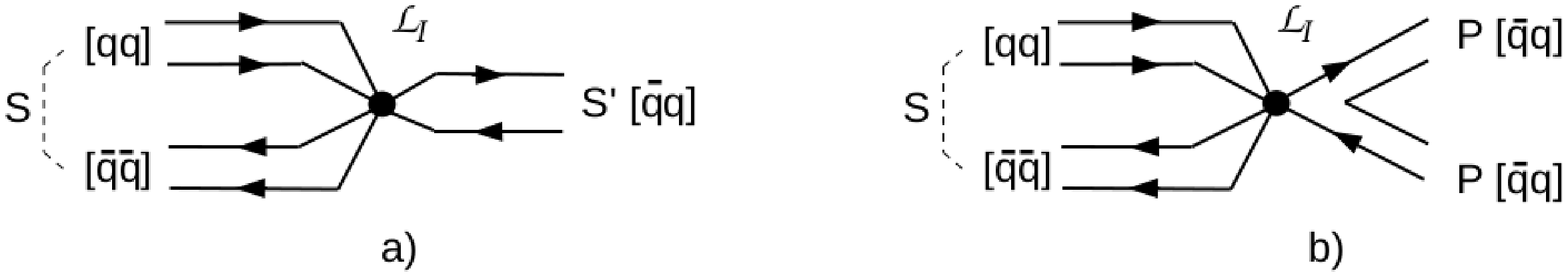}        
\caption{\footnotesize 
The two main effects of the instanton Lagrangian in the scalar sector: (a) the 
tetraquark-$q\bar q$ mixing; (b) the Zweig-rule violating $S\to PP$ amplitude.}
\label{fig:inst}
\end{center}
\end{figure}

With three light quark flavors, ${\cal L}_{\rm I}$ is proportional to the product of three 
quark and three antiquark fields, antisymmetrised in flavor and color, and it includes 
a term of the type
\be
\tr(J^{[4q]} J^{[2q]} )~,
\label{inst:flavstr}
\ee
where 
\be
J^{[4q]}_{ij} =  [{\bar q} {\bar q}]_i  [ q q]_j~,
\qquad 
J^{[2q]}_{ij} =  {\bar q}_j q_i~,
\ee
and
\begin{equation}
[qq]_{i\alpha}= \epsilon_{ijk}\epsilon_{\alpha \beta\gamma} \bar q^{j\beta}_c\gamma_5 q^{k\gamma}.
\label{defdq}
\end{equation}
$[q q]_{i\alpha}$ is the spin-0 diquark operator, latin indices indicate flavor,  
greek indices stand for color and $\bar q_c$ is the charge-conjugate of the quark field.
As shown in Fig.~\ref{fig:inst}, this term can lead to 
a mixing between tetraquark and $q\bar q$ scalar states (Fig.~\ref{fig:inst}a), 
and an effective $S$(tetraquark)$\to P P$ coupling (Fig.~\ref{fig:inst}b)
which allows the ideally-mixed $f_0$ state in (\ref{qcomp})
to decay into two pions.

\subsection{The tetraquark--$q\bar q $ mixing}

The quantum numbers of $J^{[4q]}_{ij}$ and $J^{[2q]}_{ij}$ match those 
of the tetraquark and $q\bar q $ scalar states 
$S_{ij}$ and $S^\prime_{ij}$
\bea
S &=& \left[ \ba{ccc}
\frac{a^0}{\sqrt{2}}+\frac{f_0^{[0]}}{\sqrt{2}} & a^+ &  \kappa^+  \\ 
a^- & -\frac{a^0}{\sqrt{2}} + \frac{f_0^{[0]}}{\sqrt{2}} & \kappao \\ 
\kappa^- & \kappab &  \sigma^{[0]}
\ea \right]~~{\rm (}S=[qq][\bar q \bar q]{\rm )}~,
\label{eq:S4q}  \\
S^\prime &=& \left[ \ba{ccc}
\frac{a^{\prime 0}}{\sqrt{2}}+\frac{\sigma^{\prime[0]}}{\sqrt{2}} 
  & a^{\prime +} &  \kappa^{\prime +}  \\ 
a^{\prime -} & -\frac{a^{\prime 0}}{\sqrt{2}} + \frac{\sigma^{\prime [0]}}{\sqrt{2}} 
  & {\kappa}^{\prime 0} \\ 
\kappa^{\prime -} & \bar{\kappa}^{\prime 0} &  f_0^{\prime [0]}
\ea \right]~~{\rm (}S^\prime=q\bar q{\rm )}~,
\label{eq:S2q}
\eea
where the the neutral 
isoscalar states  are not necessarily mass eigenstates. 
The instanton Lagrangian thus generate a mixing term 
\be
\cL_{\rm mix}  = \gamma ~\tr(S\cdot S^\prime)~.
\label{eq:SSmix}
\ee

A mixing of this form was introduced in~\cite{Schect}
and is  essential for a consistent identification of the 
scalar mesons around 1.5 GeV in terms of  $q\bar q$ states,  
with a possible addition of a scalar glueball~\cite{glue}.
Indeed, the well identified $I=1$ and $I=1/2$ states around 1.5~GeV, 
$a_0(1450)$ and  $K^*(1430)$, also show a reversed mass ordering, 
although  smaller than in the case of the 
light states. The anomaly can be explained as a contamination 
of the inverse hierarchy of the light tetraquark states via $\cL_{\rm mix}$.
The coefficient $\gamma$ was determined phenomenologically 
from a fit to the mass spectrum~\cite{Schect}  to be:
\be
|\gamma |\simeq 0.6~{\rm GeV}^2~.
\label{mixing2}
\ee
With this value and the observed masses, the bare masses of the lightest $q\bar q$ scalars 
turn out to be slightly above 1~GeV. The result goes well with the estimate 
obtained in~\cite{eta2} from a consistent description of pseudoscalar states 
(including the $\eta^\prime$) and scalar $q\bar q$ states within a 
linear sigma model. The bare $q\bar q$ masses agree also with the natural ordering 
of P-wave states, that predicts $0^{++}$ masses to be smaller 
than $1^{++}$ and $2^{++}$ masses~\cite{Schect,pospar}.

It was observed in~\cite{pospar}   that the value of the effective coupling 
in (\ref{mixing2}) is much larger than what expected by 
usual QCD interactions for such Zweig-rule violating 
effect.  To obtain this mixing by usual QCD interactions, 
it is necessary to annihilate completely quarks and  antiquarks in the 
initial state and to produce from vacuum those of the final state. 
A strongly suppressed transition~\cite{pospar}, which instead 
is provided almost for free by the instanton Lagrangian.

A proviso concerns the hadronic matrix element of  $J^{[2q]}$.
In the fully non-relativistic approximation, one would have:
\be
\langle 0|\bar q (0) q(0) |S^\prime \rangle =\Psi (0)~,
\ee
where $\Psi(0)$ is the non-relativistic wave-function in the origin, which vanishes for P-waves. 
However, for relativistic quark fields we get a non-vanishing result, proportional 
to $v=p/E$. For QCD, Coulomb like, bound states, $v\sim \alpha_S$, and the P-wave nature 
of $S^\prime$ results only in a mild suppression. 

The instanton induced mixing could 
in principle  be determined by the following matrix element 
\be
\langle S|\cL_{\rm I}|S^\prime \rangle \propto \langle S| \tr(J^{[4q]} J^{[2q]} )|S^\prime \rangle
\approx \tr\left(\langle S|J^{[4q]}|0 \rangle \langle 0|J^{[2q]}|S^\prime \rangle \right)~.
\ee
However, at present we do not have a reliable independent information on the
matrix elements of $J^{[4q]}$ and $J^{[2q]}$ between scalar states and 
vacuum.

\begin{figure}[t]
\begin{center}
\epsfig{height=3truecm, width=12truecm,figure=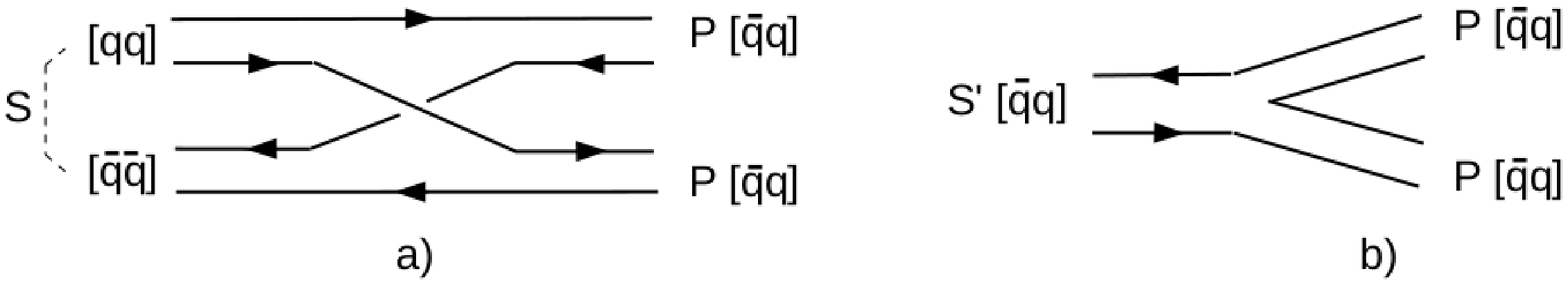}        
\caption{ \footnotesize 
Leading quark-flavor diagrams for the decays into two pseudoscalar 
mesons of tetraquark (a) and $q \bar{q}$ (b) scalar mesons.}
\label{quarkrearr}
\end{center}
\end{figure}

\subsection{$S\to PP$ decays}

The leading mechanisms describing the decays of tetraquark 
and $q\bar q$ scalar states into two pseudoscalar mesons are 
illustrated in Fig.~\ref{quarkrearr}.
In the tetraquark case, the diagram in Fig.~\ref{quarkrearr}a
(denoted as {\it quark rearrangement} in~\cite{scalar1})
explains the affinity of $f_0$ and $a_0$ to $K \bar K$ channels. 
A non vanishing $f_0\to \pi\pi$ amplitude appears if the 
possible mixing of  $f_0^{[0]}$ and $\sigma^{[0]}$ in (\ref{qcomp})
is taken into account. However, as we illustrate in a
quantitative way in the following section, 
the mixing alone does not lead to a good fit of all  $S\to PP$ 
decays~\cite{Schect,scalar1}. As already observed in~\cite{scalar1}, 
this problem indicates the presence of additional 
contributions to the $S\to PP$ amplitudes, generated by 
a different dynamical mechanism. 
This mechanism can be traced back to the 
instanton amplitude in Fig.~\ref{fig:inst}b.

In constructing effective Lagrangians for the  $S(S^\prime) \to PP$ 
decays an important role is played by chiral symmetry. 
In the chiral limit the octet components of the light pseudoscalar mesons 
\beq
\Phi = \left[ \ba{ccc}
\frac{\pi^0}{\sqrt{2}}+\frac{\eta_q}{\sqrt{2}}
& \pi^+ &  K^+ \\ 
\pi^- & -\frac{\pi^0}{\sqrt{2}} + \frac{\eta_q}{\sqrt{2}} & \Ko \\ 
 K^- & \Kob & \eta_s 
\ea \right]~, \label{eq:Phi}
\eeq
can be identified with the Goldstone bosons of the spontaneous breaking 
of ${\rm SU}(3)_L\times {\rm SU}(3)_R$ into ${\rm SU}(3)_{L+R}$. 
Following the general formalism of Ref.~\cite{CCWZ} (see also~\cite{schec2}),
the effective Lagrangian of lowest dimension allowed by 
chiral symmetry contributing to $S\to PP$ decays, 
with $P$ restricted to the octet components, 
turns out to be composed by only two independent operators:
\be
O_1(S) = \frac{F^2}{2} \tr(S u_\mu  u^\mu) \qquad {\rm and} \qquad
O_2(S) =  \frac{F^2}{2} \tr(S) \tr(u_\mu  u^\mu)~,  
\ee
where  
\be
 u_\mu  = i u\da \partial_\mu U  u\da =  -(\sqrt{2}/F) \partial_\mu \Phi + \cO(\Phi^2)~,
\qquad U = u^2 =  e^{i\sqrt{2}\Phi/F}~.
\qquad 
\ee
Being dictated only by chiral symmetry, an identical structure 
holds for the $S^\prime\to PP$  transitions of the $q\bar q$ scalar states.

The relative strength of the effective couplings of these operators 
can be determined by the correspondence of their flavor structures 
with a given quark-flavor diagram.  In the tetraquark case, the leading 
amplitude in Fig.~\ref{quarkrearr}a contributes to both $O_1$ and $O_2$, 
generating the combination $O_1 -\frac{1}{2} O_2$ \cite{scalar1}.  
Taking into account also contributions with the singlet pseudoscalar 
field, the effective operator generated is:
\be
 O_f(S) = O_1(S) -\frac{1}{2} O_2(S) + O_S(S)~, \qquad  
 O_S(S) = \frac{F^2}{2} \left[ -\tr(S u_\mu)\tr(u^\mu) + \frac{1}{2} 
 \tr(S) \tr(u_\mu)\tr(u^\mu) \right]~.
\ee

The instanton induced coupling in Fig.~\ref{fig:inst}b 
has a completely different flavor structure. From the chiral realization 
of the currents in (\ref{inst:flavstr}), 
\bea 
&&\langle PP |J^{[2q]}|0\rangle=(u_\mu u^\mu)_{\cO(\Phi^2)} + \ldots     \notag\\
&& \langle PP|\tr(J^{[4q]} J^{[2q]})|S\rangle=\tr (S u_\mu u^\mu)_{\cO(\Phi^2)}  
   + \ldots \propto \tr (S \partial_\mu \Phi \partial^\mu \Phi ) + \ldots 
\label{eq:J4SPP}
\eea
where dots denote higher-order terms in $\Phi$ and in the chiral expansion. It follows that instanton effects are encoded only by $O_1(S)$.
Taking into account both the leading quark-rearrangement diagram and 
the instanton contribution, decays of scalar tetraquark states 
into pseudoscalar mesons should  be described by the effective Lagrangian
\be
\cL_{\rm decays}(S) = c_f  O_f(S) + c_I O_1(S) 
\label{eq:lag}
\ee
where we expect $|c_I| \ll |c_f|$ since that the instanton 
contribution is a subleading effect.

Similarly to the case of the $S$--$S^\prime$ mixing, we are not able to 
evaluate the hadronic matrix element of $J^{[4q]}$ in (\ref{eq:J4SPP})
from first principles, therefore we cannot predict the value of the 
effective instanton coupling $c_I$ in (\ref{eq:lag}). 
However, an interesting crosscheck of the normalization of the instanton 
effective Lagrangians is obtained under the hypothesis 
that the leading contribution to the amplitude in Fig.~\ref{fig:inst}b 
is the $S^\prime$ pole term arising by the contraction 
of the diagrams in Fig.~\ref{fig:inst}a and Fig.~\ref{quarkrearr}b. 

The $S^\prime \to PP$ decays of the heavy $q\bar q$ states have been 
analyzed in Ref.~\cite{Schect} and found to be reasonably well described 
by the effective Lagrangian
\be
\cL_{\rm decays}(S^\prime) = c^\prime_f~O_1 (S^\prime)~,
\label{coupl2}
\ee
corresponding to the chiral realization of the diagram in 
Fig.~\ref{quarkrearr}b. The value of the effective coupling 
is found to be $c^\prime_f \approx 6.1~{\rm GeV}^{-1}$ 
($c^\prime_f=2A^\prime$ in the notation of Ref.~\cite{Schect}).
Under the plausible hypothesis of $S^\prime$ pole dominance for 
the instanton-induced $S \to PP$ amplitude we thus expect 
\be
\left| c_I^{(S^\prime {\rm -pole})} \right|
= \frac{|\gamma|~ c_f^\prime}{M^2_{S^\prime}-M^2_{S}}\sim \frac{|\gamma| 
~c_f^\prime}{M^2_{S^\prime}}\sim 0.016
\label{control}
\ee
where we have used $M_{S^\prime}\sim 1.5$ GeV. 

The above estimate should be taken cautiously.  
Non-pole terms are not expected to be totally negligible, 
and the poor experimental information about the
heavy scalar meson decays implies a sizable uncertainty in the value 
of $c^\prime_f$. Nonetheless 
it is very encouraging that the value in (\ref{control}) is
consistent with the phenomenological value of $c_I$ that we shall obtain 
in the next section from the phenomenology of
light scalar decays with the effective Lagrangian~(\ref{eq:lag}).

\section{Phenomenological analysis of $S\to PP$ decays}
\label{sect:SPP}
In this section we analyze the decays of the light scalars 
into two pseudoscalar mesons using the effective Lagrangian
in (\ref{eq:lag}). As discussed before, such Lagrangian with 
$|c_I |\ll |c_f|$ corresponds to the tetraquark hypothesis for the light 
scalar mesons, taking into account instanton effects.

The $g_{_{SP_1P_2}}$ couplings, defined by 
\beq
 {\cal A}\left( S\to P_1(p_1) P_2(p_2) \right) ~=~ g_{_{SP_1P_2}}~
p_1^\mu p_{2\mu} ~=~ g_{_{SP_1P_2}}~ \frac{1}{2} (M_S^2-M_{P_1}^2-M_{P_2}^2)~,
\eeq  
are reported in Table~\ref{tab:cap}. 
The decay rates are then given by:
\bea
&&\Gamma(S\to P_1P_2)=|{\cal A}(S\to P_1P_2)|^2\frac{p^*}{8\pi M_S^2}~, \nonumber 
\eea
where $p^*$ is the decay momentum. 

For comparison, we also attempt a fit of light 
scalar meson decays with the effective Lagrangian
\be
\cL_{\rm eff}(S^\prime)= c^\prime_f  O_1(S^\prime)+ c^\prime_I O_f(S^\prime)
\label{eq:2q}
\ee
where, contrary to what advocated so far, we identify 
the $q\bar q$ scalar field in (\ref{eq:S2q}) with the nonet 
of  light scalar mesons. In this scheme the leading operator is 
$O_1(S^\prime)$ while instanton effects contribute to $O_f(S^\prime)$ (the 
corresponding $g_{_{S P_1 P_2}}$ are reported in the third line of Table~\ref{tab:cap}).
The comparison shows that, beside the problems with the mass spectrum, 
the $q\bar q$ hypothesis for the light scalar mesons
has also serious difficulties in fitting $S\to PP$ data.

\subsection{Mass  mixing  of neutral states}
\label{sect:mixing}

\begin{table}[t]
\begin{center} 
\begin{tabular}{@{}|c|c|c|c|c|}
\hline
-- &$\sigma$ &$\kappa$ &$f_0 $ &$a_0$  \\
\hline
$M$(MeV) &  441 & 800 & 965 & 999 \\
\hline
\end{tabular}\\[2pt]
\caption{\footnotesize  Numerical values used  
for the light scalar masses (see Ref.~\cite{scalarlow,BES1,Caprini,Descotes,PDG}) }
\label{tab:mass}
\end{center}
\end{table}

The physical $f_0$ and $\sigma$ states are in general 
a superposition of the ideally mixed states $\sigma^{[0]}$ and $f^{[0]}_0$
defined in Eqs.~(\ref{eq:S4q})--(\ref{eq:S2q}).
Introducing a generic mixing between 
the mass eigenstates and the ideally mixed states
\be
\begin{bmatrix}
\sigma \\ f_0
\end{bmatrix}
=
\begin{bmatrix}
\cos\omega & \sin\omega \\ -\sin\omega & \cos\omega
\end{bmatrix}
\begin{bmatrix}
\sigma^{[0]}\\ f_0^{[0]}
\end{bmatrix}~,
\ee
the value of the mixing angle $\omega$ is determined by the experimental values of the 
scalar masses (see~\cite{scalar1,sannino}). Fixing the scalar masses to the input 
values in Table~\ref{tab:mass} but for the poorly known $\kappa$ mass, and 
letting the latter vary in the interval $[750,800]$~MeV, we find that the 
deviations from the ideal mixing case are quite small. We find in particular
$|\omega|< 5^\circ$, which we use as range to estimate the impact 
of a non-vanishing $\omega$.

Another important ingredient to compute the physical amplitudes is the 
$\eta$--$\eta^\prime$ mixing. In the octet--singlet basis we define
\bea
\begin{bmatrix}
\eta\\
\eta^\prime
\end{bmatrix}
&=&U(\phi_{_{PS}})
\begin{bmatrix}
\eta_8\\
\eta_0
\end{bmatrix},
\qquad 
\quad U(\phi_{_{PS}}) ~=~ 
\begin{bmatrix}
\cos\phi_{_{PS}} &-\sin\phi_{_{PS}}\\
\sin\phi_{_{PS}} & \cos\phi_{_{PS}}
\end{bmatrix}~.
\eea
It is useful to consider also the mixing in the quark basis:
\bea
\begin{bmatrix}
\eta\\
\eta^\prime
\end{bmatrix}
&=&U(-\theta)
\begin{bmatrix}
\eta_q\\
\eta_s
\end{bmatrix}~,
\eea
where $\phi_{_{PS}}+\theta +\tan ^{-1}(\sqrt{2})=0$.
From the analysis of the pseudoscalar meson masses, $\gamma\gamma$ decays of $\eta$ and $\eta^\prime$ and $J/\psi \to \gamma ~\eta/\eta^\prime$~\cite{ginop}, one obtains
$\phi_{_{PS}} \simeq -22^0$ ($\theta \simeq -33^0$).

The effect of $\eta-\eta^\prime$ mixing in the 
$a_0\to\eta\pi$ processes plays an important role in distinguishing 
the two hypotheses for the scalar mesons. In the four-quark case, 
the quark exchange amplitude (Fig.~\ref{quarkrearr}a) 
produces a pure $\eta_s$, while the instanton interaction
(Fig.~\ref{fig:inst}b) produces a pure $\eta_q$.
Expressing the coupling to physical particles $\eta$ and $\eta^\prime$
in terms of the octet-singlet mixing angle leads to:
\bea
&&g^{[4q]}_{a_0^+\pi^+\eta}=\sqrt{2}c_I\cos\theta - c_f \sin\theta =\frac{\sqrt{2} (c_f+c_I) \cos \phi_{_{PS}} +(c_f-2 c_I) \sin \phi_{_{PS}} }{\sqrt{3}} \label{eq:gaeta}
\eea
In the $q\bar q$ hypothesis, where the role of two operators $O_1$ and $O_f$ is exchanged, 
one finds
\bea
&&g^{[q\bar q]}_{a_0^+\pi^+\eta} = \frac{\sqrt{2} (c^\prime_f+c^\prime_I) \cos \phi_{_{PS}}
 - 
(2 c^\prime_f - c^\prime_I) \sin \phi_{_{PS}} }{\sqrt{3}} \label{eq:gaeta2}
\eea

\subsection{Numerical analysis}

\begin{table}[t]
\begin{center}
\begin{tabular}{@{}|c|c|c|c|c|c|c|c|}
\hline
&$\sigma^{[0]}\pi^+\pi^-$  & $f_0^{[0]}\pi^+\pi^-$ & $f_0^{[0]} K^+ K^-$ & $\kappa^+  K^0\pi^+$ & $a_0  \eta_q \pi$ &$a_0  \eta_s \pi$ & $a_0^-  K^- K^0$ \\ 
\hline
$[qq][\bar q\bar q]$&$-c_f$ & $\sqrt{2}c_I$  & $\frac{1}{\sqrt{2}}(-c_f+c_I)$ &$c_f+c_I$ & 
$\sqrt{2} c_I$ 
&
$- c_f $ 
& $c_f+c_I$ \\ 
\hline
$q\bar q$& $\sqrt{2}c^\prime_f$ & $-c^\prime_I$  & $c^\prime_f$ &$c^\prime_f+c^\prime_I$ & 
$\sqrt{2} c^\prime_f$ 
&
$- c^\prime_I $ 
& $c^\prime_f+c^\prime_I$ \\ 
\hline 
\end{tabular}\\[2pt]
\caption{\footnotesize The $g_{_{SP_1P_2}}$ couplings for tetraquark and $q\bar q$ 
scalar mesons, from the effective Lagrangians (\ref{eq:lag}) and (\ref{eq:2q}), 
respectively.}\label{tab:cap}
\end{center}
\end{table}

The results of the fit to $S\to PP$ amplitudes are reported in Table~\ref{tab:res}.
We use the masses given in Table~\ref{tab:mass} and $\phi_{_{PS}}=-22^\circ$~\cite{ginop}. 
We analyze both the two- and four-quark hypotheses, 
with or without the instanton contribution, using  
$\sigma\to \pi\pi$ and $f_0\to \pi\pi$ as input channels. 
For simplicity, in  Table~\ref{tab:res}
we compare the theoretical predictions with the KLOE data only.
A comparison with other experimental results can simply 
be obtained by means of Table~\ref{tab:numeriexp},
where all the available experimental information is collected.
The data for $\sigma$ and $\kappa$ decays are the results 
of the theoretical analyses in~\cite{Caprini,Descotes}. 

\begin{table}[t]
\begin{center}
\begin{tabular}{|l|ll|ll|}
\hline   
\raisebox{0pt}[10pt][3pt]{Processes}  &   \multicolumn{2}{|c|}{KLOE~\cite{KLOE1,KLOE3}}  
    & \multicolumn{2}{|c|}{BES, Crystal Barrel, WA102}    \\  \hline
\raisebox{0pt}[12pt][5pt]{$f_0 \to   \pi^+\pi^- $}  &  $1.43_{-0.60}^{+0.03}$ & $\left(1.3 \pm 0.1\right)$     
    &  $2.32 \pm 0.25$~\cite{BES}  & \\
\raisebox{0pt}[12pt][5pt]{$f_0 \to    K^+ K^- $}    &  $3.76_{-0.49}^{+1.16}$ & $\left(0.4_{-0.3}^{+0.6} \right)$ 
    &  $4.12 \pm 0.55$~\cite{BES}  & \\  \hline
\raisebox{0pt}[12pt][5pt]{$a_0 \to  \pi^0 \eta$}    &  $2.8\pm 0.1$   & $(2.2\pm 0.1)$    & $2.3 \pm 0.1$~\cite{CBall} & 
$2.1 \pm 0.23$~\cite{Barberis} \\  
\raisebox{0pt}[12pt][5pt]{$a_0 \to K^+ K^-$}        &  $2.16\pm 0.04$ & $(1.6\pm 0.1)$    & $1.6 \pm 0.3$~\cite{CBall} & \\  \hline
\end{tabular}  \\ [2pt]
\caption{\footnotesize    
Experimental data for the $S\to PP$ amplitudes in GeV. The number in brackets 
in the KLOE column refers to the parameterization of Ref.~\cite{IMNP} without the 
$\sigma$ pole (the absence of the $\sigma$ contribution makes the
$f_0 \to   \pi^+\pi^-,  K^+ K^- $ results between brackets 
less reliable for the present analysis; similarly, we do not quote the 
$f_0 \to   \pi^+\pi^-,  K^+ K^- $  results extracted from $\sigma(e^+e^- \to \pi^+\pi^-\gamma)$~\cite{KLOE2}
which suffer of a larger background). 
The numbers in the last column refer to other experiments where it has 
been possible to unambiguously extract the information on the partial amplitudes. 
\label{tab:numeriexp} }
\end{center}
\end{table}

\begin{table}[t]
\begin{center}
\begin{tabular}{|l||c|c|c|c|c|l||}
\hline
Processes  &   \multicolumn{3}{c|}{ ${\cal A}_{\rm th}([qq][\bar q \bar q])$} & \multicolumn{2}{c|}{ ${\cal A}_{\rm th}(q\bar q)$}& ${\cal A}_{\rm expt}$  \\ 
    & with inst. &  no inst. & best fit & with inst.& no inst.& \\ \hline
$\sigma   \to \pi^+\pi^- $ &   
{\rm input} &  {\rm input} & 1.6 &
{\rm input} &{\rm input}  &$3.22 \pm 0.04$   \\ \hline
$\kappa^+ \to  K^0 \pi^+  $ &   $7.3$ & 
7.7  & 3.3 & 6.0 & 5.5 &$5.2  \pm 0.1 $   \\  \hline
$f_0 \to \pi^+\pi^- $    &   {\rm input} & [0--1.6]&1.6 &
{\rm input}  & [0--1.6]&$1.4\pm 0.6$  \\  
$f_0 \to    K^+ K^- $    & $6.7$  & 6.4&3.5&  6.4 & 6.4 &$3.8\pm 1.1$  \\  \hline
$a_0 \to  \pi^0 \eta$    &  $6.7$ & 7.6 & 2.7 & 12.4 & 11.8 &$2.8\pm 0.1$    \\  
$a_0 \to K^+ K^-$        &   $4.9 $  & 5.2 & 2.2 & 4.1 & 3.7 &$2.16\pm 0.04$  \\  \hline
\end{tabular}  \\ [2pt]
\caption{\footnotesize Numerical results, amplitudes in GeV. Second and third columns: results obtained with a decay  Lagrangian including or not including instanton effects, respectively. 
Fourth column: best fit with instanton effects included (see text).
Fifth and sixth columns: predictions for a $q\bar q$ picture of the light scalars with
and without instanton contributions. All results are obtained with a $\eta$--$\eta^\prime$ mixing angle $\phi_{_{PS}}=-22^\circ$. Second and fifth columns are computed with a scalar mixing angle $\omega=0$. The $f_0\to \pi\pi$ couplings in the third and sixth columns are computed with $\omega$  in the range $\pm 5^\circ$ (see text).
Data for $\sigma$ and $\kappa$ decays are from~\cite{Caprini,Descotes}, the reported amplitudes correspond to: $\Gamma_{\rm tot}(\sigma) = 544 \pm 12$, $\Gamma_{\rm tot}(\kappa) = 557 \pm 24$.   
\label{tab:res}  }
\end{center}
\end{table}


The values of the couplings derived in the four-quark hypothesis and $\omega=0$ are: 
\be 
c_f=0.041~{\rm MeV}^{-1};~~ c_{I}=-0.0022~{\rm MeV}^{-1}~~~~(\omega=0)
\label{frominput}
\ee
The negative sign of $c_I$  is chosen to minimize the $a_0 \to \eta \pi$ rate.
As shown by  eq.~(\ref{eq:gaeta}), the negative interference
between $c_f$ and $c_{I}$ increases for $\phi_{_{PS}}$ more negative. 

As an alternative strategy, we have performed a global fit,
assigning conventionally a 10\% error to the $\sigma$ rate, 
15\% to $\kappa$, 30\%  to all others and $\phi_{_{PS}}=-22^\circ$ and searched for a best fit solution.
The results of this fit are reported in the fourth column. 
The best-fit couplings are:
\be 
c_f=0.020\pm 0.002~{\rm MeV}^{-1},~~ c_{I}=-0.0025\pm 0.0012~{\rm MeV}^{-1}~~~~(-5^\circ\leq\omega\leq 5^\circ)
\label{best-fit}
\ee
Central values are for $\omega=0$, errors are given by letting $\omega$ to 
vary in $\pm 5^\circ$ range (see sect.~\ref{sect:mixing}).

Some comments concerning the fit under the four-quark hypothesis are in order: 
\begin{itemize}
\item There is a good overall consistency, the fit is stable and, as expected, $|c_I |\ll |c_f|$.
The fitted value of $|c_I |$ is also perfectly consistent with the 
pole estimate presented in Eq.~(\ref{control}).
\item The positive feature of the instanton contribution is that it  provides a 
non-vanishing $f_0\pi\pi$ coupling independent from mixing and, at the same time, it improves the agreement with data on the `clean' $a_0\to\eta\pi$ channel. 
\item The relation between $\kappa\to K\pi$ and $a_0\to KK$ is fixed by $SU(3)$ and does 
not depend on the value of the couplings. As a result it is impossible to fit 
simultaneously the central values of these two amplitudes without 
introducing symmetry breaking terms. 
\item Being extracted from  $\sigma(e^+e^-\to \pi\pi\gamma)$ data, 
the experimental values for the $f_0,a_0\to KK$ couplings reported in 
Table~\ref{tab:res}
are subject to a sizable theoretical uncertainty (not shown in the table).
\item A mixing angle $\omega$ in the $\pm 5^\circ$ range (see sect.~\ref{sect:mixing}) affects, most significantly, the no-instanton case, leading to 
a non-vanishing $f_0\to\pi\pi$ amplitude marginally consistent with data but it does not 
solve the $a_0\to\eta\pi$ problem (Table~\ref{tab:res}, third column). 
\end{itemize}

The fit under the  $q\bar q$ hypothesis, reported in the fifth and sixth
columns of Table~\ref{tab:res}
is much worse. Due to the exchange of $f_0$ and $\sigma$ going from  $S$ to $S^\prime$, 
it remains true that there is no $f_0\to \pi\pi$ in the absence of instantons 
and, similarly to the tetraquark case,  we obtain $|c^\prime_I| \ll  |c^\prime_f|$. 
However, the situation is drastically different for the $a_0\to \pi^0\eta$ channel.
In the $q\bar q$ case there is no sign conspiracy that  produces the cancellation 
found before and the predicted amplitude is far from the observed one 
for any value of $\phi_{_{PS}}$, with or without instanton effects.
This bad fit provides a further evidence, beside the inverse mass spectrum, 
against the $q\bar q$ hypothesis for the lightest scalar mesons.

\section{Conclusions}

The addition of the instanton-induced effective  six-fermion Lagrangian 
lead us to  a simple and satisfactory description 
of both the light scalar mesons below 1 GeV 
and the heavier scalar states around 1.5 GeV. The light mesons are 
predominantly tetraquark states ($S$), while the heavier ones are predominantly
$q{\bar q}$ states ($S^\prime$). Instantons induce a mixing between the two 
sets of states, which  explains the
puzzling mass spectrum of the heavier mesons. 
Moreover, integrating out the heavy states, instanton effects  
manifest themselves in the dynamics of the lightest states, 
generating the Zweing-rule  violating  amplitude which 
is necessary for a consistent description of 
the strong decays of the light scalar mesons.

The phenomenological determinations of the $S$--$S^\prime$ mixing 
and of the  Zweig-rule  violating  $S\to PP$ amplitude suggest
that the dynamics of the scalar mesons could be described, 
to a good extent, by a simple effective Lagrangian of the type
\be
\cL_{\rm eff,all} = \tr\left( S{\cal M}_S^2 S\right)+\tr\left(S^\prime {\cal M}^2_{S^\prime} 
S^\prime\right)+\gamma ~\tr\left( S S^\prime \right)+ c_f O_f (S) + c^\prime_f O_1 (S^\prime)
\label{eq:ltot}
\ee
where ${\cal M}^2_{S^{(\prime)}}=a^{(\prime)} + b^{(\prime)} \lambda_8$ 
are appropriate mass matrices reflecting the inverse and normal mass 
ordering of tetraquark and $q{\bar q}$ states, 
and $\gamma$ is the coupling encoding the instanton contribution.


\paragraph{Note Added} After this work was submitted, an
independent analysis of the instanton-induced mixing
between two and four quark scalar mesons has been presented in~\cite{Schechternew}.

\paragraph{Acknowledgements}
We thank Prof. A. Zichichi for the exciting environment provided by the Erice School of Subnuclear Physics, 
where this work was initiated. We thank Gene Golowich for his critical comments on the manuscript.

\end{document}